\documentclass[vecphys]{svmult}

\usepackage{makeidx}        
\usepackage{graphicx}       
                            
\usepackage{multicol}       
\usepackage[bottom]{footmisc}

\makeindex

\def\be{\begin{equation}}
\def\ee{\end{equation}}
\def\hompc{\,h\,{\rm Mpc}^{-1}}
\def\bd{\mbox{\boldmath $\delta$}}
\def\C{\mbox{\boldmath $C$}}
\def\x{\mbox{\boldmath $x$}}
\def\xp{\mbox{\boldmath $x'$}}
\def\k{\mbox{\boldmath $k$}}
\def\llangle{\left\langle}
\def\rrangle{\right\rangle}
\def\kms{\,{\rm km\,s^{-1}}}

\begin{document}

\title*{Cosmological constraints from galaxy clustering}
\author{Will J. Percival\inst{1}}
\institute{Institute of Cosmology and Gravitation, University of
Portsmouth, Portsmouth, P01 2EG. \texttt{will.percival@port.ac.uk}}
\maketitle

\section{Abstract}

In this manuscript I review the mathematics and physics that
underpins recent work using the clustering of galaxies to derive
cosmological model constraints. I start by describing the basic
concepts, and gradually move on to some of the complexities involved
in analysing galaxy redshift surveys, focusing on the 2dF Galaxy
Redshift Survey (2dFGRS) and the Sloan Digital Sky survey
(SDSS). Difficulties within such an analysis, particularly dealing
with redshift space distortions and galaxy bias are highlighted. I
then describe current observations of the CMB fluctuation power
spectrum, and consider the importance of measurements of the
clustering of galaxies in light of recent experiments. Finally, I
provide an example joint analysis of the latest CMB and large-scale
structure data, leading to a set of parameter constraints.

\section{introduction}

The basic techniques required to analyse galaxy clustering were
introduced in the 70s \cite{peebles73}, and have been subsequently
refined to match data sets of increasing quality and size. In this
manuscript I have tried to summarise the current state of this
field. Obviously, such an attempt can never be complete or unique in
every detail, although it is still worthwhile as it is always useful
to have more than one source of information. An excellent alternative
viewpoint was recently provided by Hamilton
\cite{hamilton1,hamilton2}, which covers some of the same material,
and provides a more detailed review of some of the statistical methods
that are used. Additionally it is worth directing the interested
reader to a number of good text books that cover this topic
\cite{coles_lucchin,dodelson,liddle_lyth,book_of_john}. In addition to
a description of the basic mathematics and physics behind a clustering
analysis I have attempted to provide a discussion of some of the
fundamental and practical difficulties involved. The cosmological goal
of such an analysis is consider in the final part of this manuscript,
where the combination of cosmological constraints from galaxy
clustering and the CMB is discussed, and an example multi-parameter
fit to recent data is considered.

\section{Basics}
\label{sec:basics}

Our first step is to define the dimensionless overdensity
\be
  \delta({\bf x}) = \frac{\rho({\bf x})-\bar{\rho}}{\bar{\rho}},
\ee
where $\bar{\rho}$ is the expected mean density, which is independent
of position because of statistical homogeneity.

The autocorrelation function of the overdensity field (usually just
referred to as the correlation function) is defined as
\be
  \xi(\mathbf{x_1},\mathbf{x_2})
    \equiv \llangle
    \delta(\mathbf{x_1})\delta(\mathbf{x_2})
    \rrangle.
\ee
From statistical homogeneity and isotropy, we have that
\begin{eqnarray}
  \xi(\mathbf{x_1},\mathbf{x_2})
    & = & \xi(\mathbf{x_1}-\mathbf{x_2}), \\
    & = & \xi(|\mathbf{x_1}-\mathbf{x_2}|).
\end{eqnarray}
To help to understand the correlation function, suppose that we have
two small regions $\delta V_1$ and $\delta V_2$ separated by a
distance $r$. Then the expected number of pairs of galaxies with one
galaxy in $\delta V_1$ and the other in $\delta V_2$ is given by
\be
  \llangle n_{\rm pair} \rrangle = \bar{n}^2\left[1+\xi(r)\right]
    \delta V_1\delta V_2,
  \label{eq:npair}
\ee 
where $\bar{n}$ is the mean number of galaxies per unit volume. We see
that $\xi(r)$ measures the excess clustering of galaxies at a
separation $r$. If $\xi(r)=0$, the galaxies are unclustered (randomly
distributed) on this scale -- the number of pairs is just the expected
number of galaxies in $\delta V_1$ times the expected number in
$\delta V_2$. $\xi(r)>0$ corresponds to strong clustering, and
$\xi(r)<0$ to anti-clustering. Estimation of $\xi(r)$ from a sample of
galaxies will be discussed in Section~\ref{sec:xi_est}.

It is often convenient to consider perturbations in Fourier space. In
cosmology the following Fourier transform convention is most commonly used 
\begin{eqnarray}
  \delta({\bf k}) & \equiv & 
    \int\delta({\bf r})e^{i{\bf k.r}}d^3r
    \label{eq:delr}\\  
  \delta({\bf r}) & = & 
    \int\delta({\bf k})e^{-i{\bf k.r}}
    \frac{d^3k}{(2\pi)^3}.
\end{eqnarray}
The power spectrum is defined as 
\be
  P({\bf k_1},{\bf k_2})=\frac{1}{(2\pi)^3}
    \llangle\delta({\bf k_1})\delta({\bf k_2})\rrangle.
\ee
Statistical homogeneity and isotropy gives that
\be
  P({\bf k_1},{\bf k_2})=\delta_D({\bf k_1}-{\bf k_2})P(k_1),
\ee
where $\delta_D$ is the Dirac delta function. The power spectrum is
sometimes presented in dimensionless form
\be
  \Delta^2(k) = \frac{k^3}{2\pi^2}P(k).
\ee
The correlation function and power spectrum form a Fourier pair
\begin{eqnarray}
  P(k) & \equiv & 
    \int\xi(r)e^{i{\bf k.r}}d^3r\\  
  \xi(r) & = & 
    \int P(k)e^{-i{\bf k.r}}
    \frac{d^3k}{(2\pi)^3},
\end{eqnarray}
so they provide the same information. The choice of which to use is
therefore somewhat arbitrary (see \cite{hamilton1} for a further
discussion of this).

The extension of the 2-pt statistics, the power spectrum and
the correlation function, to higher orders is straightforward with
Eq.~\ref{eq:npair} becoming
\be
  \llangle n_{\rm tuple} \rrangle = \bar{n}^n\left[1+\xi^{(n)}\right]
    \delta V_1 \cdot\cdot\cdot \delta V_n.
\ee
However, the central limit theorem implies that a density distribution
is asymptotically Gaussian in the limit where the density results from
the average of many independent processes. The overdensity field has
zero mean by definition, so is completely characterised by either the
correlation function or the power spectrum. Consequently, in this
regime, measuring either the correlation function or the power
spectrum provides a statistically complete description of the field.

\section{matter perturbations}
\label{sec:seed}

\begin{figure}
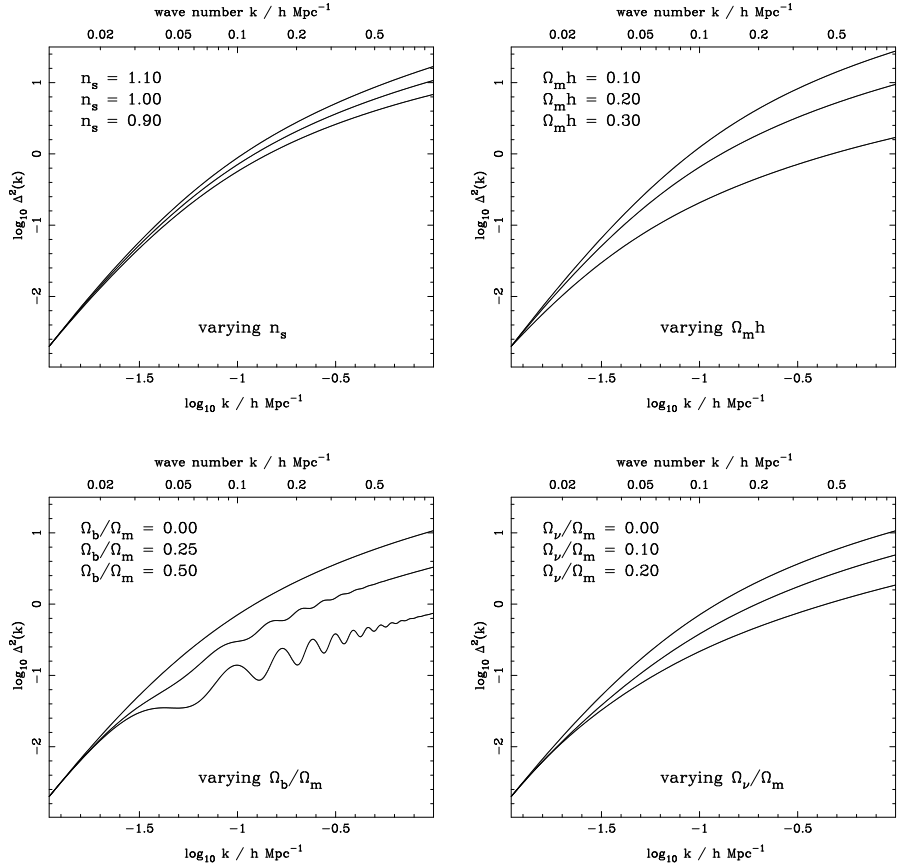

  \centering

  \includegraphics[width=0.48\textwidth]{linpk_varyn.ps}
  \hfill
  \includegraphics[width=0.48\textwidth]{linpk_varyommh.ps}

  \vspace{5mm}
  \includegraphics[width=0.48\textwidth]{linpk_varybfrac.ps}
  \hfill
  \includegraphics[width=0.48\textwidth]{linpk_varynufrac.ps}

  \caption{Plots showing the linear power spectrum (solid lines) for a
  variety of different cosmological parameters. Only the shapes of the
  power spectra are compared, and the amplitudes are matched to the
  same large scale value. Our base model has $\Omega_Mh=0.2$, $n_s=1$,
  $\Omega_b/\Omega_M=0$ and $\Omega_\nu/\Omega_M=0$. Deviations from
  this base model are given in each panel. As can be seen many of the
  shape distortions from changing different parameters are similar,
  which can cause degeneracies between these parameters when fitting
  models to observations.}
  \label{fig:linpk}
\end{figure}

There are three physical stages in the creation and evolution of
perturbations in the matter distribution. First, primordial
perturbation are produced in an inflationary epoch. Second, the
different forms of matter within the Universe affect these primordial
perturbations. Third, gravitational collapse leads to the growth of
these fluctuations. In this section we will discuss the form of the
perturbations on scales where gravitational collapse can be described
by a linear change in the overdensity. The gravitational collapse of
perturbations will be considered in Section~\ref{sec:pert_evol}.

\subsection{why are there matter perturbations?}

A period of ``faster than light'' expansion in the very early Universe
solves a number of problems with standard cosmology. In particular, it
allows distant regions that appear causally disconnected to have been
connected in the past and therefore explains the flatness of the
CMB. Additionally it drives the energy density of the Universe close
to the critical value and, most importantly for our discussion of
perturbations, it provides a mechanism for producing seed
perturbations as quantum fluctuations in the matter density are
increased to significant levels. For a detailed examination of the
creation of fluctuations see \cite{liddle93}. For now, we will just
comment that the most basic inflationary models give a spectrum of
fluctuations $P(k)\propto k^n$ with $n\sim1$.

\subsection{the effect of dark matter}

The growth of dark matter fluctuations is intimately linked to the
Jeans scale. Perturbations smaller than the Jeans scale do not collapse
due to pressure support -- for collision-less dark matter this is
support from internal random velocities. Perturbations larger than the
Jeans scale grow through gravity at the same rate, independent of
scale. In a Universe with just dark matter and radiation, the Jeans
scale grows to the size of the horizon at matter-radiation equality,
and then reduces to zero when the matter dominates. We therefore see
that the horizon scale at matter-radiation equality will be imprinted
in the distribution of fluctuations -- this scale marks a turn-over in
the growth rate of fluctuations. What this means in practice is that
there is a cut-off in the power spectrum on small scales, dependent on
$\Omega_Mh$, with a stronger cut-off predicted for lower $\Omega_Mh$
values. This is demonstrated in Fig.~\ref{fig:linpk}.

\subsection{the effect of baryons}

At early epochs baryons are coupled to the photons and, if we consider
a single fluctuation, a spherical shell of gas and photons is driven
away from the perturbation by a sound wave. When the photons and gas
decouple, a spherical shell of baryons is left around a central
concentration of dark matter. As the perturbation evolves through
gravity, the density profiles of the baryons and dark matter grow
together, and the perturbation is left with a small increase in
density at a location corresponding to the sound horizon at the end of
the Compton drag epoch \cite{bashinsky1,bashinsky2}. This real-space
``shell'' is equivalent to oscillations in the power spectrum. In
addition to these acoustic oscillations, fluctuations smaller than the
Jeans scale, which tracks the sound horizon until decoupling, do not
grow, while large fluctuations are unaffected and continue to
grow. The presence of baryons therefore also leads to a reduction in
the amplitude of small scale fluctuations. For more information and
fitting formulae for the different processes a good starting point is
\cite{eh98}.

\subsection{the effect of neutrinos}
\label{sec:pert_nu}

The same principal of gravitational collapse versus pressure support
can be applied in the case of massive neutrinos. Initially the
neutrinos are relativistic and their Jeans scale grows with the
horizon. As their temperature decreases their momenta drop, they
become non-relativistic, and the Jeans scale decreases -- they can
subsequently fall into perturbations. Massive neutrinos are
interesting because even at low redshifts the Jeans scale is
cosmologically relevant. Consequently the linear power spectrum (the
fluctuation distribution excluding the non-linear collapse of
perturbations) is not frozen shortly after matter-radiation
equality. Instead its form is still changing at low
redshifts. Additionally, the growth rate depends on the scale - it is
suppressed until neutrinos collapse into perturbations, simply because
the perturbations have lower amplitude. The effect of neutrino mass on
the present day linear power spectrum is shown in
Fig.~\ref{fig:linpk}. Note that in this plot the relative amplitudes
of the power spectra have been removed - it is just the shape that is
compared. The amplitude would also depend on the combined neutrino
mass.

\section{the evolution of perturbations}
\label{sec:pert_evol}

Having discussed the form of the linear perturbations, we will now
consider how perturbations evolve through gravity in the matter and
dark energy dominated regimes. To do this, we will use the spherical
top-hat collapse model, where we compare a sphere of background
material with radius $a$, with one of radius $a_p$ which contains the
same mass, but has a homogeneous change in overdensity. The ease with
which the behaviour can be modeled follows from Birkhoff's theorem,
which states that a spherically symmetric gravitational field in empty
space is static and is always described by the Schwarzchild metric
\cite{birkhoff23}. This gives that the behaviour of the homogeneous
sphere of uniform density and the background can be modeled using the
same equations. For simplicity we initially only consider the sphere
of background material.

The sphere of background material behaves according to the standard
Friedmann and cosmology equations
\be
  E^2(a) 
    = \frac{1}{a^2}\left(\frac{da}{dH_0t}\right)^2
    = \Omega_Ma^{-3} + \Omega_Ka^{-2} + \Omega_Xa^{f(a)},
\ee
\be
  \frac{1}{a}\frac{d^2a}{dt^2}
    = -\frac{H_0^2}{2}\left[\Omega_Ma^{-3} 
      + [1+3w(a)]\Omega_Xa^{f(a)}\right].
\ee
These equations have been written in a form allowing for a general
time-dependent equation of state for the dark energy
$p=w(a)\rho$. Conservation of energy for the dark energy component
provides the form of $f(a)$
\be
  f(a) = \frac{-3}{\ln a}\int_0^{\ln a}\left[1+w(a')\right]d\ln a'.
\ee

\begin{figure}
  \centering
  \includegraphics[width=9cm]{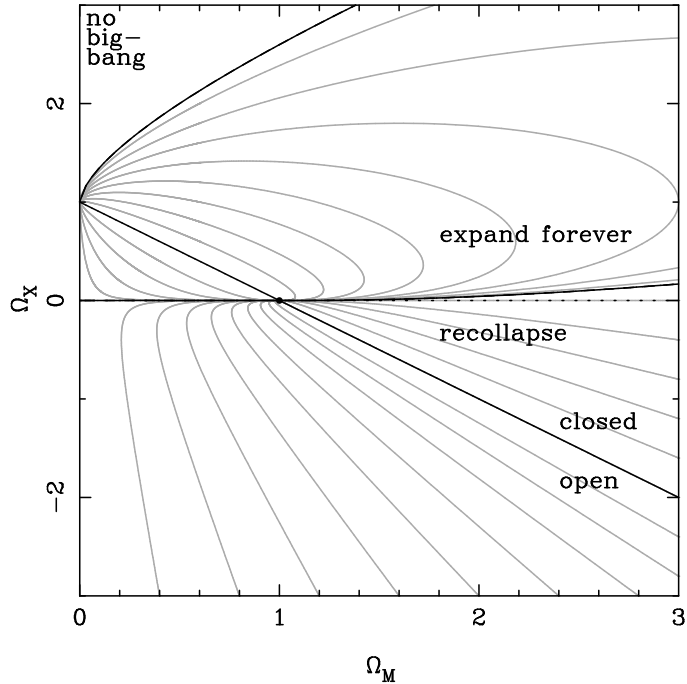}
  \caption{Plot showing the evolution of the matter and vacuum energy
  densities for a selection of cosmologies (grey lines) with constant
  dark energy equation of state parameter $w=-1$. The critical models
  that border the different types of evolution are shown by the black
  lines. The dotted line highlights $\Omega_X=0$.}
  \label{fig:energy_universe}
\end{figure}

The dark matter and dark energy densities evolve according to 
\be
  \Omega_M(a)=\frac{\Omega_Ma^{-3}}{E^2(a)},
    \,\,\, 
  \Omega_X(a)=\frac{\Omega_Xa^{f(a)}}{E^2(a)}.
\ee
Tracks showing the evolution of $\Omega_M(a)$ and $\Omega_X(a)$ are
presented in Fig.~\ref{fig:energy_universe} for $h=0.7$ and constant
dark energy equation of state $w=-1$.
Of particular interest are solutions which predict recollapse, but
that have $\Omega_X>0$. Provided that $\Omega_M>>\Omega_X$, the
perturbation will collapse before the dark energy dominates. For a
cosmology with $\Omega_M\sim0.3$ and $\Omega_X\sim0.7$, these
solutions correspond to overdense spheres that will collapse and form
structure.

For the perturbation, the cosmology equation can be written
\be
  \frac{1}{a_p}\frac{d^2a_p}{dt^2}
    = -\frac{H_0^2}{2}\left[\Omega_Ma_p^{-3} 
      + [1+3w(a)]\Omega_Xa^{f(a)}\right],
  \label{eq:cosmo_pert}
\ee 
where it is worth noting that the dark energy component is dependent
on $a$ rather than $a_p$. This does not matter for
$\Lambda$-cosmologies as $f(a)=0$, and the $a$ dependence in this term
is removed. For other dark energy models, this dependence follows if
the dark energy does not cluster on the scales of interest. For such
cosmological models, we cannot write down a Friedmann equation for the
perturbation because energy is not conserved \cite{wang98}. We also
have to be more careful using virialisation arguments to analyse the
behaviour of perturbations \cite{percival05}.

To first order, the overdensity of the perturbation
$\delta=a^3/a_p^3-1$ evolves according to
\be
  \frac{d^2\delta}{d(H_0t)^2}
    + \frac{2}{a}\frac{da}{d(H_0t)}\frac{d\delta}{d(H_0t)}
    - \frac{3}{2}\Omega_Ma^{-3}\delta = 0,
  \label{eq:lintheory}
\ee
which is known as the linear growth equation.

\begin{figure}
  \centering
  \includegraphics[width=9cm]{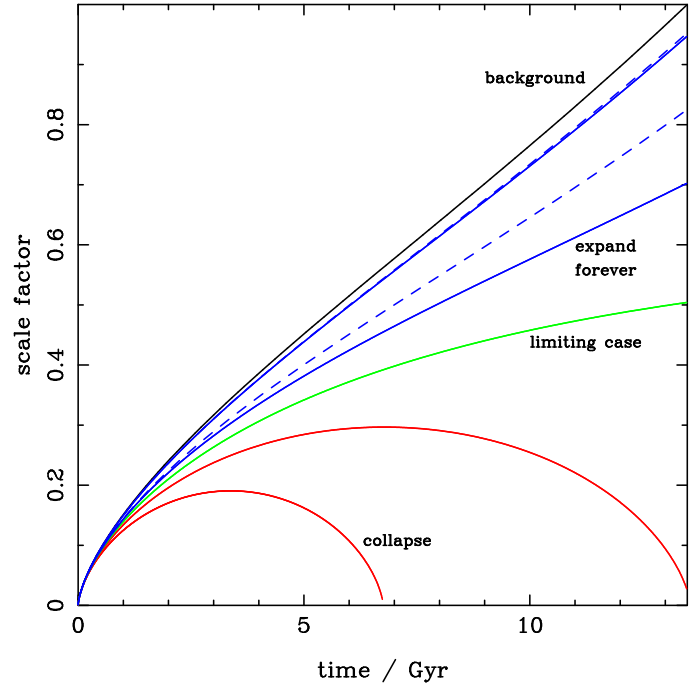}
  \caption{Plot showing the evolution of the scale factor of
  perturbations with different initial overdensities. A standard
  cosmology with $\Omega_M=0.3$, $\Omega_X=0.7$, $h=0.7$, $w=-1$ is
  assumed. The dashed lines show the linear extrapolation of the
  perturbation scales for the two least overdense perturbations.}
  \label{fig:pert_evol}
\end{figure}

The evolution of the scale factor of the perturbations is given by the
solid lines in Fig.~\ref{fig:pert_evol}, compared with the background
evolution for a cosmology with $\Omega_M=0.3$, $\Omega_X=0.7$,
$h=0.6$, $w=-1$. These data were calculated by numerically solving
Eq.~\ref{eq:cosmo_pert}. For comparison, the dashed lines were
calculated by extrapolating the initial perturbation scales using the
linear growth factor, calculated from Eq.~\ref{eq:lintheory}. Dashed
lines are only plotted for the two least overdense perturbations. In
comparison, the most overdense perturbations are predicted to collapse
to singularities. However, in practice inhomogeneities, and the
non-circular shape of actual perturbations will mean that the object
virialises with finite extent.

\begin{figure}
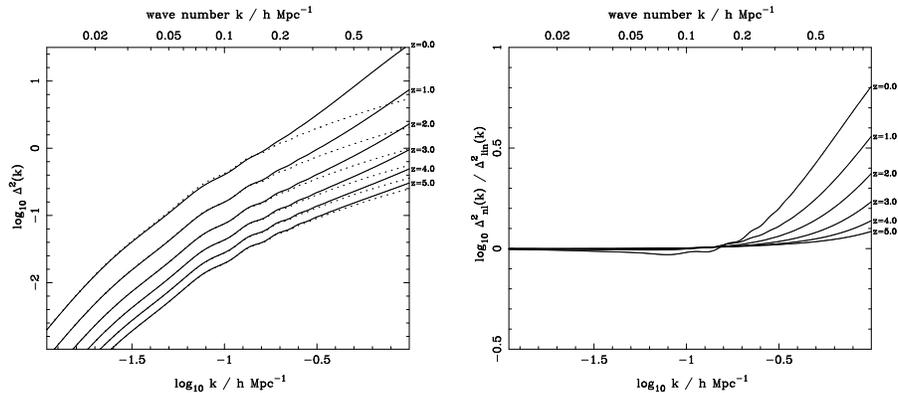

  \centering
  \includegraphics[width=0.48\textwidth]{nlpk_std.ps}
  \hfill
  \includegraphics[width=0.48\textwidth]{nlpk_rat.ps}
  \caption{Plots comparing non-linear (solid lines) and linear power
  spectra (dotted lines) at a series of redshifts from $z=0$ to
  $z=5$. In the left panel the raw dimensionless power spectra are
  plotted while in the right panel the ratio between non-linear and
  linear predictions is shown. As can be seen, on large scales linear
  growth simply increases the amplitude of the power spectrum, while
  on small scales we also see an increase in power as structures
  collapse at low redshifts. There is also a slight decrease in power
  on intermediate scales -- it is this power that is transferred to
  small scales. Non-linear power spectra were calculated from the
  fitting formulae of \cite{smith03} with $\Omega_M=0.3$, $h=0.7$,
  $n_s=1$, and $\Omega_b/\Omega_m=0.15$.}
  \label{fig:nlpow}
\end{figure}

The evolution of perturbations has a profound affect on the present
day power spectrum of the matter fluctuations on small scales. On the
largest scales, the overdensities are small and linear theory
(Eq.~\ref{eq:lintheory}) holds. This increases the amplitude of the
fluctuations, but does not change the shape of the power spectrum, as
the perturbation all grow at the same rate (except if neutrinos are
cosmologically relevant -- see Section~\ref{sec:pert_nu}). However, on
the smallest scales, overdensities are large and collapse to virialised
structures (e.g. cluster of galaxies). The effect on the power
spectrum is most easily quantified using numerical simulations, and
power spectra calculated from fitting formulae derived from such
simulations \cite{smith03} are plotted in Fig.~\ref{fig:nlpow}.

\section{galaxy survey analysis}
\label{sec:galsurvey}

\subsection{estimating the correlation function}
\label{sec:xi_est}

First suppose that we have a single population of objects forming a
Poisson sampling of the field that we wish to constrain. This is too
simple an assumption for the analysis of modern galaxy redshift
surveys, but it will form a starting point for the development of the
analysis tools required.

First we define the (unweighted) galaxy density field
\be
  n_g({\bf r}) \equiv \sum_i\delta_D({\bf r}-{\bf r_i}).
  \label{eq:ng}
\ee
The definition of the correlation function then gives
\be
  \llangle n_g({\bf r})n_g({\bf r'}) \rrangle
    = \bar{n}({\bf r})\bar{n}({\bf r'})
    [1+\xi({\bf r}-{\bf r'})]
    + \bar{n}({\bf r})\delta_D({\bf r}-{\bf r'}).
  \label{eq:xi_def}
\ee
The final term in this equation relates to the shot noise, and only
occurs for zero separation so can be easily dealt with.

In order to estimate the correlation function, we can consider a
series of bins in galaxy separation and make use of
Eq.~\ref{eq:xi_def}. Suppose that we have created a (much larger)
random distribution of points that form a Poisson sampling of the
volume occupied by the galaxies, then
\be
  1+\xi = 
    \frac{\llangle DD\rrangle}{\llangle RR\rrangle}
    (1+\xi_\Omega),
  \label{eq:xi_obs}
\ee
where $DD$ is the number of galaxy-galaxy pairs within our bin in
galaxy separation divided by the maximum possible number of
galaxy-galaxy pairs (ie. for $n$ galaxies the maximum number of
distinct pairs is $n(n-1)/2$). Similarly $RR$ is the normalised number
of random-random pairs, and we can also define $DR$ as the normalised
number of galaxy-random pairs.

If the true mean density of galaxies $\bar{n}({\bf r})$ is estimated
from the sample itself (as is almost always the case), we must include
a factor $(1+\xi_\Omega)$ that corrects for the systematic offset
induced. $\xi_\Omega$ is the mean of the two-point correlation
function over the sampling geometry \cite{landy93}. Given only a
single clustered sample it is obviously difficult to determine
$\xi_\Omega$, and the integral constraint (as it is known) remains a
serious drawback to the determination of the correlation function from
small samples of galaxies.

Because the galaxy and random catalogues are uncorrelated, $\llangle
DR\rrangle=\llangle RR\rrangle$, and we can consider a number of
alternatives to Eq.~\ref{eq:xi_obs}. In particular
\be
  1+\xi = 
    \left(1+\frac{\llangle(D-R)^2\rrangle}
                 {\llangle RR\rrangle}\right)
    (1+\xi_\Omega),
\ee
has been shown to have good statistical properties \cite{landy93}.

\subsection{estimating the power spectrum}
\label{sec:pk_est}

In this section we consider estimating the power spectrum by simply
taking a Fourier transform of the overdensity field
\cite{baumgart91,FKP,PVP}. As for our estimation of the correlation
function, suppose that we have quantified the volume occupied by the
galaxies by creating a large random catalogue matching the spatial
distribution of the galaxies, but with no clustering (containing
$\alpha$ times as many objects). The (unnormalised) overdensity field
is
\be
  F({\bf r}) = n_g({\bf r}) - n_r({\bf r})/\alpha,
\ee
where $n_g$ is given by Eq.~\ref{eq:ng}, and $n_r$ is similarly
defined for the random catalogue.

Taking the Fourier transform of this field, and calculating the power
gives
\be
  \llangle |F({\bf k})|^2 \rrangle
    = \int \frac{d^3k'}{(2\pi)^3} 
    [P({\bf k'})-P(0)\delta_D({\bf k})]
    |G({\bf k}-{\bf k'})|^2
  + (1+\frac{1}{\alpha})\int d^3r \bar{n}({\bf r}),
  \label{eq:fkp_pk}
\ee
where $G({\bf k})$ if the Fourier transform of the window function,
defined by 
\be
  G({\bf k}) \equiv \int\bar{n}({\bf r})e^{i{\bf k.r}}d^3r,
\ee 
and the final term in Eq.~\ref{eq:fkp_pk} gives the shot noise. In
contrast to the correlation function, there is a shot noise
contribution at every scale. The integral constraint has reduced to
subtracting a single Dirac delta function from the centre of the
unconvolved power - as before this allows for the fact that we
do not know the mean density of galaxies.

\subsection{complications}

There are two complications which constitute the main hindrance to
using clustering in galaxy surveys to constrain cosmology. They are
redshift space distortions -- systematic deviations in measured
redshift in addition to the Hubble flow, and galaxy bias -- the fact
that galaxies do not form a Poisson sampling of the underlying matter
distribution. Denoting the measurement of a quantity in redshift space
(galaxy distances calculated from redshifts) by a superscript $^s$ and
in real space (true galaxy distances) by $^r$, we can write the
measured power spectrum $P_{\rm gal}^s$ as
\be
  {P_{\rm gal}^s\over P_{\rm mass}} =
    {P_{\rm gal}^s\over P_{\rm gal}^r} \times
    {P_{\rm gal}^r\over P_{\rm mass}}.
\ee
The first of these terms corresponds to redshift space distortions,
while the second corresponds to galaxy bias.

\subsubsection{redshift space distortions}
\label{sec:zspace_dist}

There are two key mechanisms that systematically distort galaxy
redshifts from their Hubble flow values. First, structures are
continually growing through gravity, and galaxies fall into larger
structures. The infall velocity adds to the redshift, making the
distance estimates using the Hubble flow wrong. This means that
clusters of galaxies appear thinner along the line-of-sight, causing
an increase in the measured power. In the distant observer
approximation, the apparent amplitude of the linear density
disturbance can be readily calculated \cite{kaiser87}, leading to a
change in the power corresponding to
\be
  P_{\rm gal}^s = P_{\rm gal}^r (1+\beta \mu^2)^2,
\ee
where $\beta=\Omega_M^{0.6}/b$, $b$ is an assumed linear bias for the
galaxies, and $\mu$ is the cosine between the velocity vector and the
line-of-sight. In the small angle approximation, we average over a
uniform distribution for $\mu$ giving
\be
   P_{\rm gal}^s = P_{\rm gal}^r
    \left[1+\frac{2}{3}\beta+\frac{1}{5}\beta^2\right].
\ee
For large redshift surveys of the nearby Universe, the small angle
approximation breaks down, although a linear result can be obtained
using a spherical expansion of the survey (see
Section~\ref{sec:sphbases}).

When objects collapse and virialise they attain a 
distribution with some velocity dispersion. These random velocities
smear out the collapsed object along the line of sight in redshift
space, leading to the existence of linear structures 
pointing towards the observer. These structures, known as
``fingers-of-god'' can be corrected by matching with a group catalogue
and applying a correction to the galaxy field before analysis
\cite{tegmark04}. Alternatively, if the pairwise distribution of
velocity differences is approximated by an exponential distribution,
then   
\be
  P_{\rm gal}^s = P_{\rm gal}^r (1+k^2\mu^2\sigma_p^2/2)^{-1},
\ee
where $\sigma_p\sim400\kms$ is the pairwise velocity dispersion 
\cite{hawkins03}.

\subsubsection{galaxy bias}

By the simple phrase ``galaxy bias'' astronomers quantify the
``messy'' astrophysics of galaxy formation. It is common to assume a
local linear bias with $\delta_{\rm gal}=b\delta_{\rm mass}$, which
leads to a simple relation between power spectra $P_{\rm gal}^r=b^2
P_{\rm mass}$. If this bias is independent of the scale probed, then
there is nothing to worry about -- the galaxy and matter power spectra
have the same shape. However, it is well known that galaxies of
different types have different clustering strengths -- two recent
analyses are \cite{seaborne99,wild05}.

One simple way of understanding galaxy bias is to use the ``halo
model'', which has become popular over the last 5 years
\cite{seljak00,peacock00,cooray02}. First, consider the distribution
of the underlying matter -- the power spectrum was shown in
Fig.~\ref{fig:nlpow}. There are two distinct regimes: on large scales,
linear growth holds, while on small scales the dark matter has formed
into halos: it has either undergone collapse and has virialised, or is
on the way to virialisation. Galaxies pinpoint certain locations
within the dark matter halos, according to an occupation distribution
for each galaxy type. This forms a natural environment in which to
model galaxy bias, with galaxies of different luminosities and types
have different occupation distributions depending on the physics of
their formation.

For 2-pt statistics, then there are two possibilities for pairs of
galaxies. We could have chosen a pair where both galaxies lie in the
same halo -- this is most likely on small scales. Alternatively, the
galaxies might be in different halos -- this is most likely on large
scales. On large scales, the halos themselves are biased compared with
the matter and we can use the peak-background split model
\cite{cole89,mo96,sheth99} to estimate the increase in clustering
strength. This limiting large scale value offers a route to determine
the masses of the virialised structures in which particular galaxies
live.

Given a linear bias model for each type of galaxy in the sample to be
analysed, it is possible to multiply the contribution of each galaxy
to the estimate of the overdensity field by the inverse of an expected
bias \cite{PVP}. Provided the bias model is correct (and possibly
altered for each scale observed), then this removes any systematic
offset in the recovered power spectrum caused by galaxy bias. The
problem is that we need to have an accurate model of the galaxy bias
in order to remove it.

\subsection{weights}

The procedure described in Section~\ref{sec:pk_est} can be extended to
include weights for each galaxy in order to optimise the analysis
\cite{FKP}. Under the assumptions that the wavelength of interest
$2\pi/k$ is small compared with the survey scale (i.e. the window is
negligible), and that the fluctuations are Gaussian, then the optimal
weight applied to galaxy $i$ is
\be
  w_i = \frac{1}{1+\bar{n}({\bf r}_i)\hat{P}(k)},
\ee
where $\bar{n}({\bf r}_i)$ is the mean galaxy density at the location
of galaxy $i$. At locations where the mean galaxy density is low,
galaxies are weighted equally. Where the galaxy density is
high, we weight by volume. It is worth noting that the optimal weights
also depend on an estimate of the power spectrum to be measured, and
therefore depend on the scale of interest. However, in practice this
dependence is sufficiently weak that very little information is lost
by assuming a constant $\hat{P}(k)$.

It is possible to include galaxy bias when determining weights and
optimising the analysis in order to recover the most signal. Given a
bias for each galaxy $b_i$ (which can be dependent on any galaxy
properties and the scale of interest), then the optimal weighting is
\cite{PVP}.
\be
  w_i = \frac{b_i^2}
    {1+\sum_j\bar{n}({\bf r_i},b_j)b_j^2\hat{P}(k)},
\ee
which up-weights the most biased galaxies that contain the strongest
cosmological signal.

\subsection{spherical bases}
\label{sec:sphbases}

\begin{figure}
  \centering
  \includegraphics[width=0.24\textwidth]{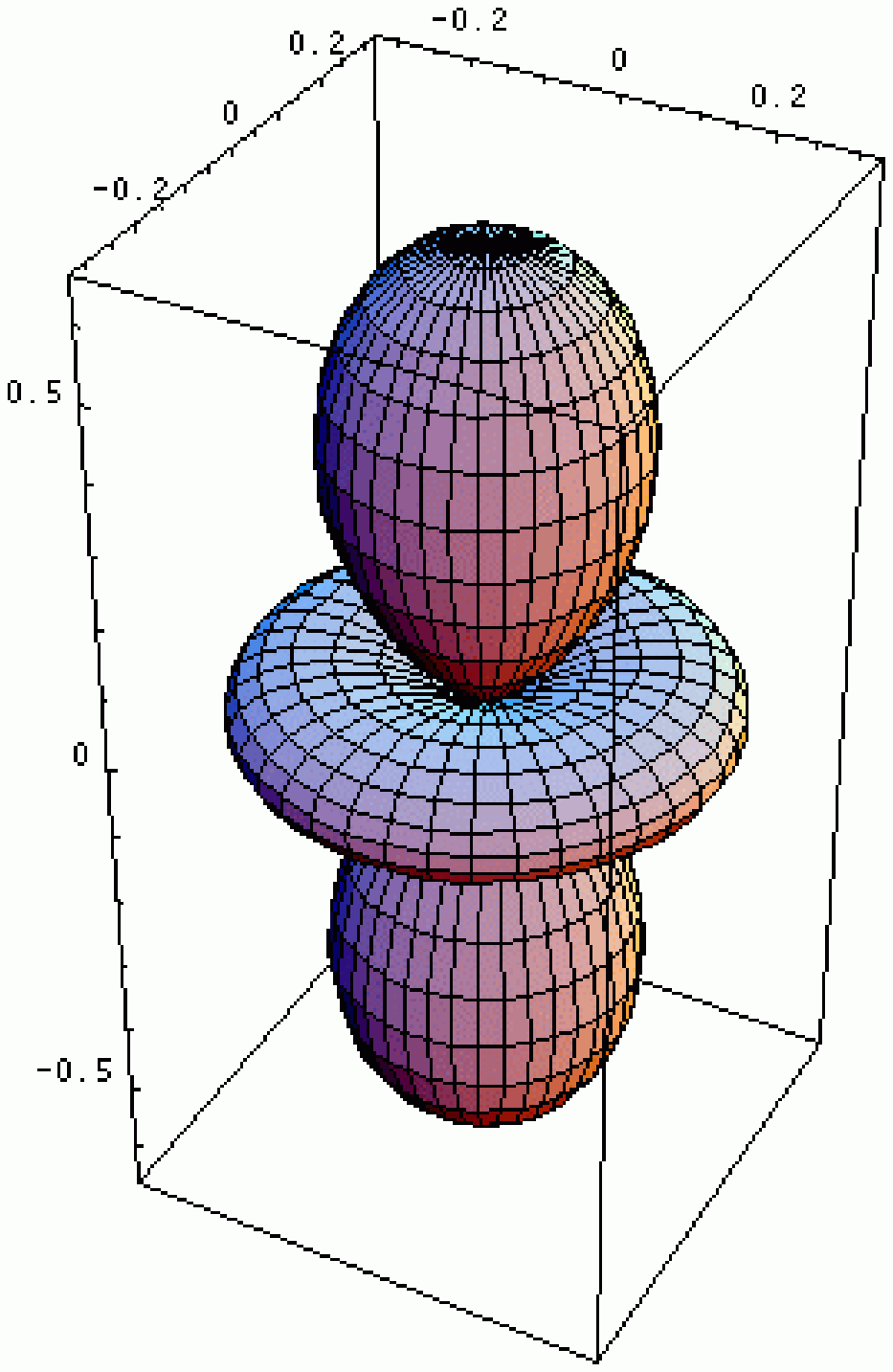}
  \includegraphics[width=0.24\textwidth]{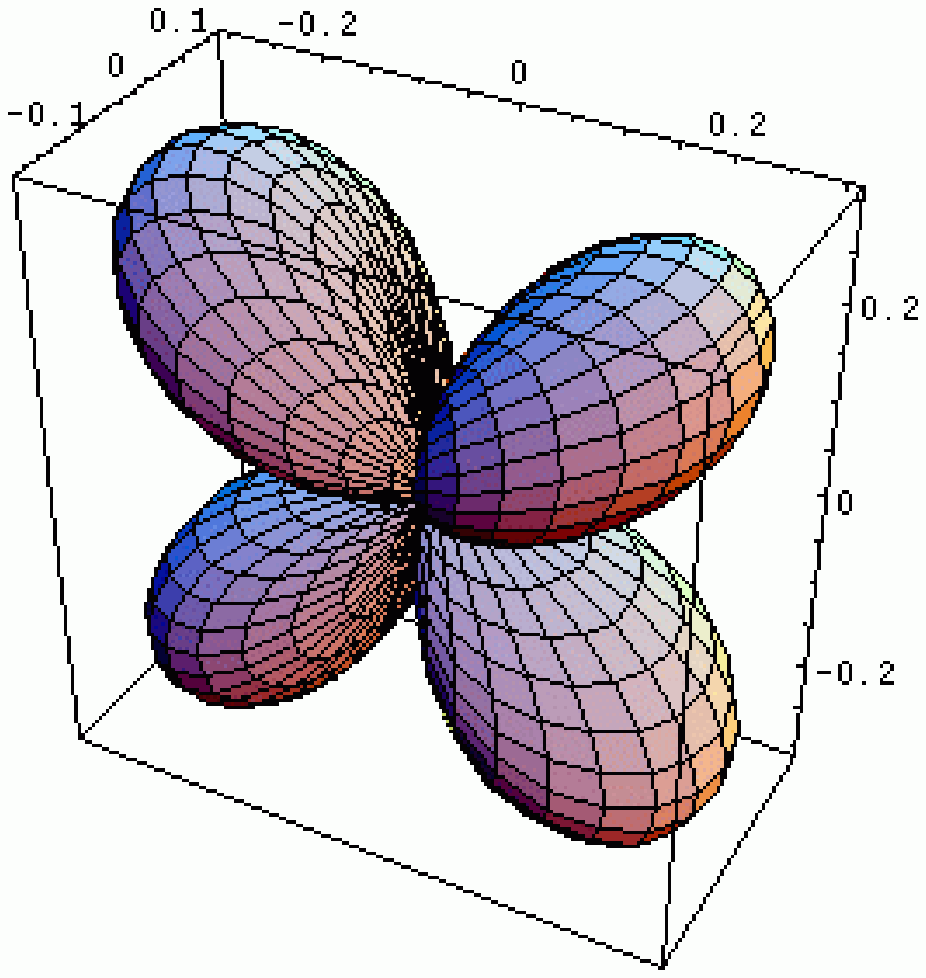}
  \hfill
  \includegraphics[width=0.48\textwidth]{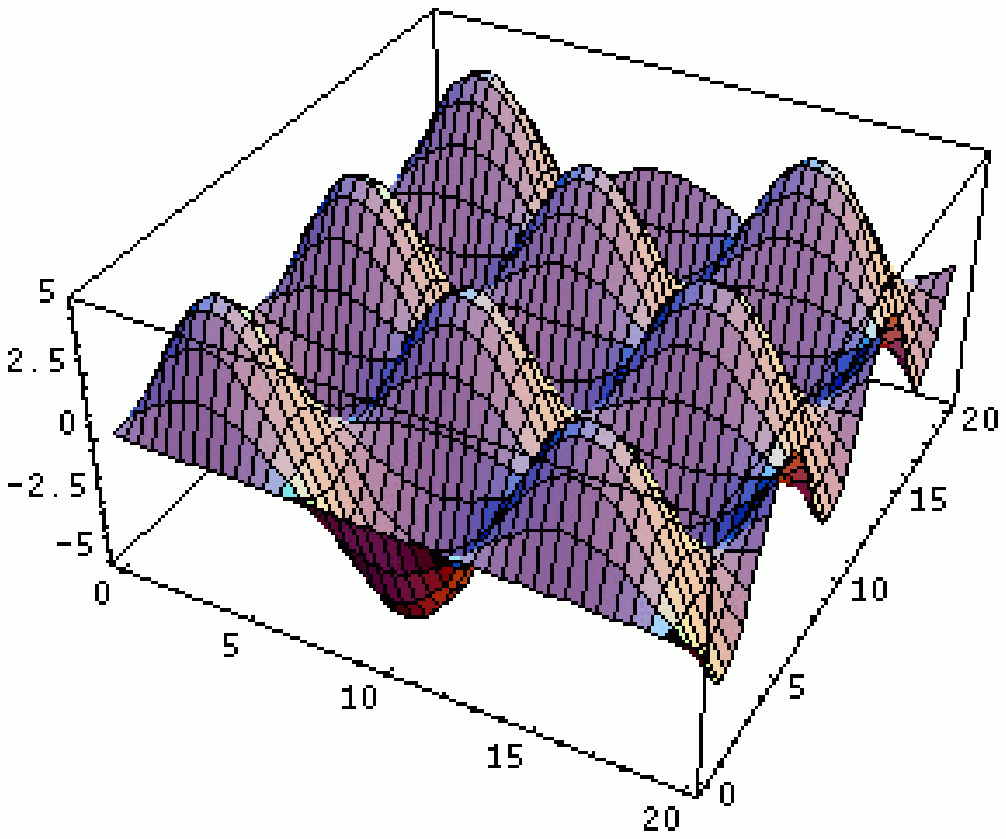}

  \vspace{1mm}  
  \includegraphics[width=0.48\textwidth]{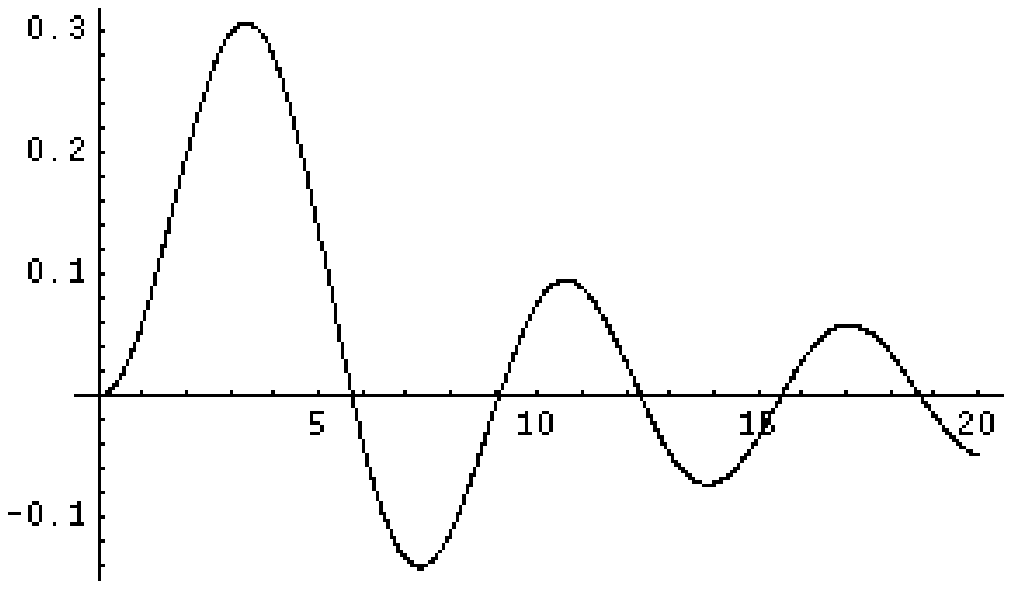}
  \hfill
  \includegraphics[width=0.48\textwidth]{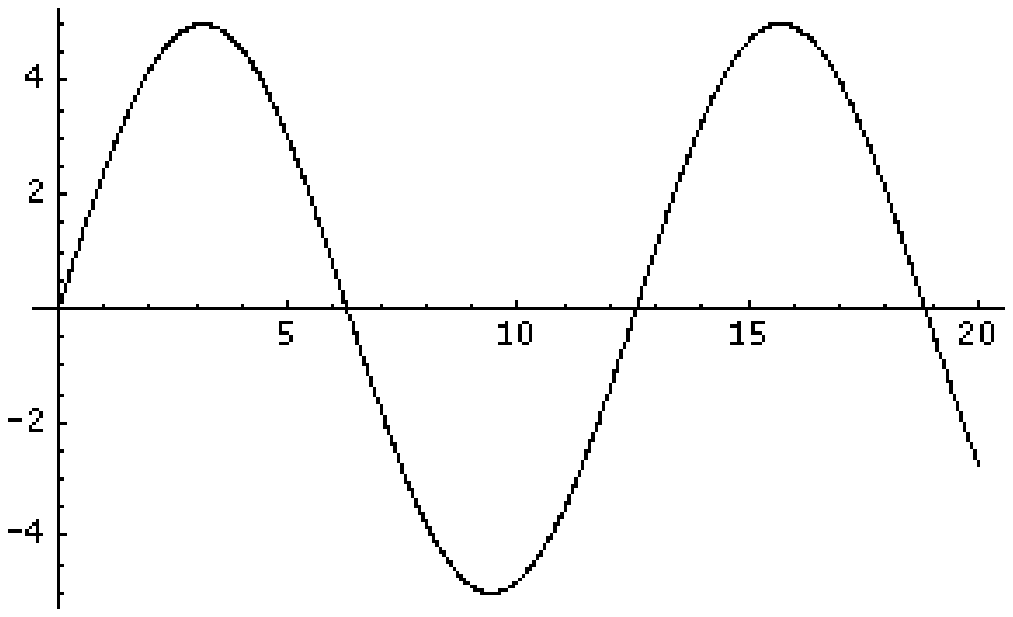}
 
  \caption{Comparison of 3D Fourier basis split into 2D and 1D
  components (right) with basis of Spherical Harmonics (with $l=2$ and
  $m=0,1$ -- top left) and Spherical Bessel functions (bottom left).}
  \label{fig:bases}
\end{figure}

In Section~\ref{sec:pk_est} we described the most simple analysis
method for a 3-dimensional galaxy survey -- decomposing into a 3D
Fourier basis. However, as we discussed in
Section~\ref{sec:zspace_dist} redshift-space distortions complicate
the situation, and cannot easily be dealt with using a Fourier basis. By
decomposing into a basis that is separable in radial and angular
directions, we can more easily correct such distortions. A pictorial
comparison of the Fourier basis with a radial-angular separable basis
is presented in Fig.~\ref{fig:bases}.

In this section we provide an overview of a formalism to do this based
on work by \cite{heavens95,tadros99,percival04}. For alternative
formalisms see \cite{fisher94,hamilton2,tegmark04}. In comparison with
the Fourier decomposition (Eq.~\ref{eq:delr}), we decompose into a 3D
basis of Spherical Harmonics $Y_{lm}$ and spherical Bessel functions
$j_l$
\be
  \delta(\x)=\sqrt{\frac{2}{\pi}}\int^\infty_0\sum_{l,m}
    \delta_{lm}(k)j_l(kx)Y_{lm}(\theta,\phi)kdk.
\ee
Because of the choice of bases, the transformation
$\delta_{lm}(k)\leftrightarrow k\delta(\k)$ is unitary so we retain the
benefit of working with the Fourier power spectrum
\be
  \llangle \delta_{lm}(k) \delta_{l'm'}(k') \rrangle
    = P(k) \delta_D(k-k')\delta_D(l-l')\delta_D(m-m').
\ee
As in Section~\ref{sec:pk_est}, we have simplified the analysis by not
including any galaxy weights, although these can be introduced into
the formalism.  Additionally, it is easier to work with a fixed
boundary condition - usually that fluctuations vanish at some large
radius so that we are only concerned with radial modes that have
\be
  \left.\frac{d}{dx}j_l(kx)\right|_{x_{\rm max}} = 0,
\ee
so that the decomposition becomes
\be
  \delta(\x)=\sum_{l,m,n} c_{ln}
    \delta_{lmn}j_l(k_{ln}x)Y_{lm}(\theta,\phi),
\ee
where $c_{ln}$ is a normalising constant.

In order to analyse the transformed modes, we need a model for
$\llangle\delta_{lmn}\delta_{l'm'n'}\rrangle$. First we deal with the
survey volume by introducing a convolution
\be
  \hat{\delta}_{lmn}=\sum_{l'm'n'} M_{lmn}^{l'm'n'} \delta_{l'm'n'},
\ee
where
\be
  M_{lmn}^{l'm'n'} = c_{ln} c_{l'n'} \int\!d^3\!\x \bar{\rho}(\x) 
    j_{l}(k_{ln}x) j_{l'}(k_{l'n'}x) 
    Y^*_{lm}(\theta,\phi) Y_{l'm'}(\theta,\phi).
\ee
We can include the effect of linear redshift space distortions by a
transform
\be
  j_l(k_{ln}x^s) \simeq j_l(k_{ln}x^r) + 
    \Delta x_{\rm lin} \frac{d}{dx^r}j_l(k_{ln}x^r),
\ee
where
\be
  \Delta x_{\rm lin} = \beta \sum_{lmn} \frac{1}{k_{ln}^2} c_{ln}
    \delta_{lmn} \frac{dj_l(k_{ln}x^r)}{dx^r} Y_{lm}(\theta,\phi).
\ee
Here $\beta=\Omega_M^{0.6}/b$. The bias $b$ corrects for the fact that
while we measure the galaxy power spectrum, the redshift space
distortions depend on the mass. We can also introduce a further
convolution to correct for the small-scale fingers-of-god effect
\be
  \hat{\delta}_{l'm'n'}=\sum_{l''m''n''} 
    S_{l'm'n'}^{l''m''n''} \delta_{l''m''n''},
\ee
where
\be
  S_{l'm'n'}^{l''m''n''} = c_{l'n'} c_{l''n''} 
    \delta^D_{l'l''} \delta^D_{m'm''}
    \int\int p(r-y)j_{l'}(k_{l'n'}r)j_{l''}(k_{l''n''}y)\,r\,dr\,y\,dy,  
\ee
and $p(r-y)$ is the 1-dimensional scattering probability for the
velocity dispersion. It is also possible to include bias and evolution
corrections in the analysis method \cite{percival04}.

For a given cosmological model, we can use the above formalism to
calculate the covariance matrix
$\llangle\delta_{lmn}\delta_{l'm'n'}\rrangle$ for $N$ modes, and then
calculate the Likelihood of a given cosmological model assuming that
$\hat{\bd}_{lmn}$ has a Gaussian distribution
\be
  {\cal L}[\hat{\bd}_{lmn}|{\rm model}] = 
    \frac{1}{(2 \pi)^{N/2} |{\C}|^{1/2}}
    {\rm exp}\left[ - \frac{1}{2} 
    \hat{\bd}_{lmn}^T \C^{-1} \hat{\bd}_{lmn}\right],  
\ee
where $C$ is the matrix of $\llangle\delta_{lmn}\delta_{l'm'n'}\rrangle$.

\section{practicalities}
\label{sec:practicalities}

\subsection{brief description of redshift surveys}
 
The 2dF Galaxy Redshift Survey (2dFGRS), which is now complete, covers
approximately 1800 square degrees distributed between two broad
strips, one across the South Galactic pole and the other close to the
North Galactic Pole, plus a set of 99 random 2~degree fields spread
over the full southern galactic cap.  The final catalogue contains
reliable redshifts for 221\,414 galaxies selected to an
extinction-corrected magnitude limit of approximately $b_J=19.45$
\cite{colless03}.

In contrast, the Sloan Digital Sky Survey (SDSS) is an ongoing
photometric and spectroscopic survey. The SDSS includes two
spectroscopic galaxy surveys: the main galaxy sample which is complete
to a reddening-corrected Petrosian $r$ magnitude brighter than
$17.77$, and a deeper sample of luminous red galaxy sample selected
based on both colour and magnitude \cite{eisenstein01}. The SDSS has
regular public data releases: the 4th data release in 2005 included
480000 independent galaxy spectra \cite{sdssdr4}. When completed, the
SDSS will have obtained spectra for $\sim10^6$ galaxies.

\subsection{angular mask}

\begin{figure}
  \centering
  \includegraphics[width=9cm]{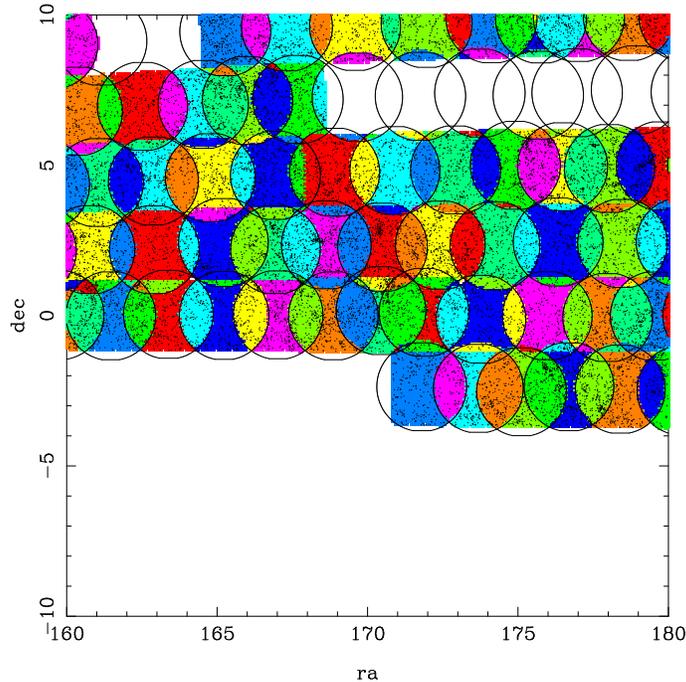}
  \caption{Section in the SDSS DR4 angular mask showing the positions
  of galaxies with measured redshifts (black dots), the positions of
  the plates from which the spectra were obtained (large black
  circles) and the segments within the mask that have different
  completenesses (coloured regions).}
  \label{fig:sdss_mask_segment}
\end{figure}

Both the recent 2dF galaxy redshift (2dFGRS) and the ongoing Sloan
Digital Sky Survey (SDSS) adopted an adaptive tiling system in order
to target photometrically selected galaxies for spectroscopic
follow-up. The circular tiles within which spectra could be taken in a
single pointing of the telescope were adaptively fitted over the
survey region, with regions of high galaxy density being covered by
two or more tiles. A region of such tiling is shown in
Fig.~\ref{fig:sdss_mask_segment}. This procedure divides the survey
into segments, each with a different completeness - the ratio of good
quality spectra to galaxies targeted. It is usually assumed that this
completeness is uniform across each of the segments formed by
overlapping tiles. Understanding this completeness is a major
consideration when performing a large-scale structure analysis of
either of these surveys. Note that the distribution of segments
depends on all adjoining targeted tiles, not just those that have been
observed.

As well as understanding the completeness, we also need to consider
the effect of the weather - spectra taken under bad observing
conditions will tend to preferentially give redshifts for nearby
rather than distant galaxies. We also need to worry about bad fields -
regions near bright stars where photometric data is of poor
quality. For the SDSS, there are hard limits for the spectroscopic
region depending on how much photometric data was available when the
targeting algorithm was run. All of these effects are well known and
can be included in an analysis.

\subsection{radial distribution}

\begin{figure}
  \centering
  \includegraphics[width=7cm]{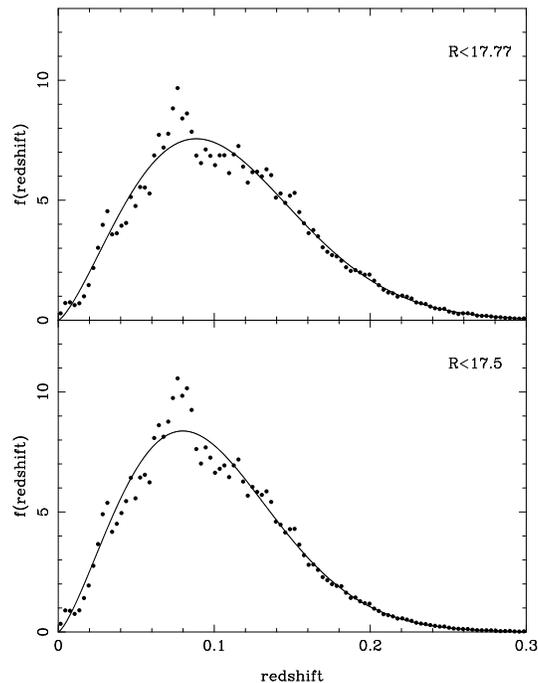}
  \caption{Redshift distribution of spectroscopically observed
  galaxies within the SDSS DR4 with apparent R magnitude less than
  $17.5$ and $17.77$ (solid circles). For comparison we show the best
  fit model given by Eq.~\ref{eq:zfit} for each distribution (solid
  lines).}
  \label{fig:sdss_redshift_distribution}
\end{figure}

In addition to the angular distribution of galaxies, we also need to
be able to model the radial distribution -- in the formalism
introduced in Section~\ref{sec:pk_est}, we need this information in
order to create the random catalogue. Perhaps the best way of doing
this is to model the true luminosity function of the distribution of
observed galaxies, and then apply a magnitude cut-off. This was the
procedure adopted in \cite{cole05}. However, the reduction in the
amplitude of the recovered power spectrum caused by fitting to the
redshift distribution is small and it is common to simply fit a
functional form to the distribution. In
Fig.~\ref{fig:sdss_redshift_distribution} we present the distribution
of galaxy redshifts in the SDSS DR4 sample compared with a fit of the
form \cite{baugh93}
\be
  f(z) = z^g \exp\left[-\left(\frac{z}{z_c}\right)^b\right],
  \label{eq:zfit}
\ee
where $g$, $b$ and $z_c$ are free parameters that have been fitted to
the data.

\section{results from recent surveys}
\label{sec:interpreting}

\subsection{results}

In Table~\ref{tab:lss_constraints} we summarise recent cosmological
constraints derived from the 2dFGRS and SDSS. In order to provide a
fair test of different analyses, we have only presented best-fit
parameters and errors for $\Omega_Mh$, fixing the other important
parameters. Degeneracies between parameters, caused by the similarity
between power spectrum shapes shown in Fig.~\ref{fig:linpk} mean that,
it is only the most recent analyses of the largest samples that can
simultaneously constrain 2 or more of these parameters.  In
Table~\ref{tab:lss_constraints} we also presented the number of galaxy
redshifts used in each analysis.

\begin{table}
  \centering
  \caption{Summary of recent cosmological constraints from 2dFGRS and
  SDSS galaxy redshift surveys. To try to provide a fair comparison,
  we only present the best-fit value and quoted error for $\Omega_Mh$
  assuming that all other cosmological parameters are fixed ($n_s=1$,
  $h=0.72$, $\Omega_b /\Omega_M =0.17$, $\Omega_\nu /\Omega_M =0.0$),
  and marginalise over the normalisation.}
  \label{tab:lss_constraints}
  \begin{tabular}{lcclc}
    \hline\noalign{\smallskip}
    survey & reference\,\, & \,\,galaxy redshifts\,\, & 
    \,\,method & \,\,$\Omega_Mh$ \\
    \noalign{\smallskip}\hline\noalign{\smallskip}
    2dFGRS   & \cite{percival01}   & 166490 & Fourier & 
      $0.206\pm0.023$ \\
    2dFGRS   & \cite{percival04}   & 142756 & Spherical Harmonics\,\, & 
      $0.215\pm0.035$ \\
    2dFGRS   & \cite{cole05}       & 221414 & Fourier &  
      $0.172\pm0.014$ \\
    SDSS     & \cite{pope04}       & 205484 & KL analysis & 
      $0.207\pm0.030$ \\
    SDSS     & \cite{tegmark04}    & 205443 & Spherical Harmonics & 
      $0.225\pm0.040$ \\
    SDSS LRG & \cite{eisenstein05} & 46748  & correlation function & 
      $0.185\pm0.015$ \\
    \noalign{\smallskip}\hline
  \end{tabular}
\end{table}

\begin{figure}
  \centering
  \includegraphics[width=\textwidth]{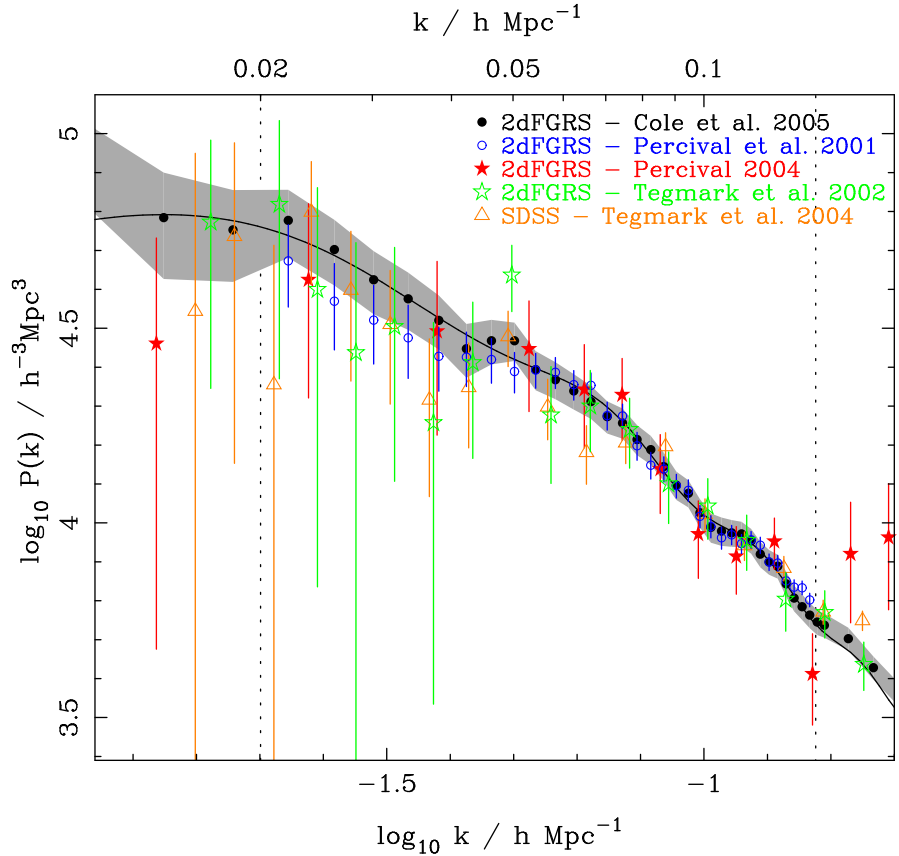}
  \caption{Plot comparing galaxy power spectra calculated by different
  analysis techniques for different surveys. The redshift-space power
  spectrum calculated by \cite{cole05} (solid circles with 1-$\sigma$
  errors shown by the shaded region) are compared with other
  measurements of the 2dFGRS power spectrum shape by \cite{percival01}
  -- open circles, \cite{percival04} -- solid stars, \cite{tegmark02}
  -- open stars. Where appropriate the data have been corrected to
  remove effects of the survey volume, by calculating the effect on a
  model power spectrum with $\Omega_Mh=0.168$, $\Omega_b /\Omega_M
  =0.0$, $h=0.72$ \& $n_s=1$. A zero-baryon model was chosen in order
  to avoid adding features into the power spectra. All of the data
  are renormalized to match the power spectrum of \cite{cole05}. The
  open triangles show the uncorrelated SDSS real space $P(k)$ estimate
  of \cite{tegmark04}, calculated using their `modeling method' with
  no FOG compression (their Table~3). These data have been corrected
  for the SDSS window as described above for the 2dFGRS data. The
  solid line shows a model linear power spectrum with
  $\Omega_Mh=0.168$, $\Omega_b /\Omega_M =0.17$, $h=0.72$, $n_s=1$ and
  normalization matched to the 2dFGRS power spectrum.}
  \label{fig:pk_comparison}
\end{figure}

The power spectra recovered from these analyses are compared in
Fig.~\ref{fig:pk_comparison}. We have corrected each for survey window
function effects using the best-fit model power spectrum. The
amplitudes have also been matched, so this plot merely shows the
shapes of the spectra. It is clear that the general shape of the
galaxy power spectrum is now well known, and the turn-over is detected
at high significance. The exact position of the turn-over is however,
more poorly known and by examining the final column of
Table~\ref{tab:lss_constraints}, we see that there are discrepancies
between recent analyses at the $\sim2\sigma$ level.

\section{combination with CMB data}

In this section we consider recent CMB observations and see how the
complementarity between CMB and large scale structure constraints can
break degeneracies inherent in these data. The major steps required
in a joint analysis are described, leading up to
Section~\ref{sec:final_results}, in which we present the constraints
from an example fit to recent data.

\subsection{cosmological models}
\label{sec:model_param}

Before we start looking at constraining cosmological models using CMB
and galaxy $P(k)$ data, it is worth briefly introducing the set of
commonly used cosmological parameters (for further discussion see the
recent review by \cite{lahav06}). It is standard to assume Gaussian,
adiabatic fluctuations, and we will not discuss alternatives here. It
is possible to parameterise the cosmological model using a number of
related sets of parameters. It is vital in any analysis that the model
that is being fitted to the data is fully specified -- including
parameters and assumed priors. Many parameters have values that
simplify the theory from which the models are calculated (e.g. the
assumption that the total density in the Universe is equal to the
critical density). Whether the data justify dropping one of these
assumptions is an interesting Bayesian question \cite{liddle04}, which
is outside the remit of the overview presented here, and we will
simply introduce the parameters commonly used and possible assumptions
about their values.

First, we need to know the geometry of the Universe, parameterised by
total energy density $\Omega_{\rm tot}$, or the curvature $\Omega_K$,
with the ``simplified'' value being that the energy density is equal
to the critical value ($\Omega_{\rm tot}=1$, $\Omega_K=0$). We also
need to know the constituents of the energy density, which we
parameterise by the dark matter density $\Omega_c$ , baryon density
$\Omega_b$, and neutrino density $\Omega_\nu$. Although it is commonly
assumed that the combined neutrinos mass has negligible cosmological
effect. The combined matter density
$\Omega_M=\Omega_c+\Omega_b+\Omega_\nu$ could also be defined as a
parameter, replacing one of the other density measurements. We also
need to specify the dark energy properties, particularly the equation
of state $w(a)$, which is commonly assumed to be constant $w(a)=-1$,
so this field is equivalent to $\Lambda$. The perturbations after
inflation are specified by the scalar spectral index $n_s$, with
$n_s=1$ being the most simple assumption. Possible running of this
spectral index is parameterised by $\alpha=dn_s/dk$ if included. A
possible tensor contribution parameterised by the tensor spectral
index $n_t$, and tensor-to-scalar ratio $r$ is sometimes explicitly
included. The evolution to present day is parameterised by the Hubble
constant $h$, and for the CMB the optical depth to last-scattering
surface $\tau$. Finally, three parameters that are often ignored and
marginalised over are the galaxy bias $b(k)$ (often assumed to be
constant) and the CMB beam $B$ and calibration $C$ errors.

\subsection{the MCMC technique}

Large multi-parameter likelihood calculations are computationally
expensive using grid-based techniques. Consequently, the Markov-Chain
Monte-Carlo (MCMC) technique is commonly used for such analyses. While
there is publically available code to calculate cosmological model
constraints \cite{lewis02}, the basic method is extremely simple and
relatively straightforward to code.

The MCMC method provides a mechanism to generate a random sequence of
parameter values whose distribution matches the posterior probability
distribution of a Bayesian analysis. Chains are sequentially
calculated using the Metropolis algorithm \cite{metropolis53}: given a
chain at position $\x$, a candidate point $\xp$ is chosen at random
from a proposal distribution $f(\xp|\x)$. This point is always
accepted, and the chain moves to point $\xp$, if the new position has
a higher likelihood. If the new position $\xp$ is less likely than
$\x$, then $\xp$ is accepted, and the chain moves to point $\xp$ with
probability given by the ratio of the likelihood of $\xp$ and the
likelihood of $\x$. In the limit of an infinite number of steps, the
chains will reach a converged distribution where the distribution of
chain links are representative of the likelihood hyper-surface, given
any symmetric proposal distribution $f(\xp|\x)=f(\x|\xp)$ (the Ergodic
theorem: see, for example, \cite{roberts96}).

It is common to implement dynamic optimisation of the sampling of the
likelihood surface (see \cite{gilks96} for examples). Again, it is
simple to assume a multi-variate Gaussian proposal function, centered
on the current chain position. Given such a proposal distribution, and
an estimate of the covariance matrix for the likelihood surface at
each step, the optimal approach for a Gaussian likelihood would
proceed as follows.

Along each principal direction corresponding to an eigenvector of the
covariance matrix, the variance $\sigma^2$ of the multi-variate
Gaussian proposal function should be set to be a fixed multiple of the
corresponding eigenvalue of the covariance matrix. To see the
reasoning behind this, consider translating from the original 17
parameters to the set of parameters given by the decomposition along
the principal directions of the covariance matrix each divided by the
standard deviation in that direction. In this basis, the likelihood
function is isotropic and the parameters are uncorrelated. Clearly an
optimized proposal function will be the same in each direction, and we
have adjusted the proposal function to have precisely this
property. There is just a single parameter left to optimize -- we are
free to multiply the width of the proposal function by a constant in
all directions. But we know that the optimal fraction of candidate
positions that are accepted should be $\sim0.25$ \cite{gelman96}, so
we can adjust the normalization of the proposal width to give this
acceptance fraction. Note that the dynamic changing of the proposal
function width violates the symmetry of the proposal distribution
$f(\xp|\x)$ assumed in the Metropolis algorithm. However, this is not
a problem if we only use sections of the chains where variations
between estimates of the covariance matrix are small.

The remaining issue is convergence -- how do we know when we have
sufficiently long chains that we have adequately sampled the posterior
probability. A number of tests are available \cite{gelman92,verde03},
although it's always a good idea to perform a number of sanity checks
as well -- for example, do we get the same result from different
chains started a widely separated locations in parameter space?

\subsection{introduction to the CMB}

\begin{figure}
  \centering
  \includegraphics[width=7cm]{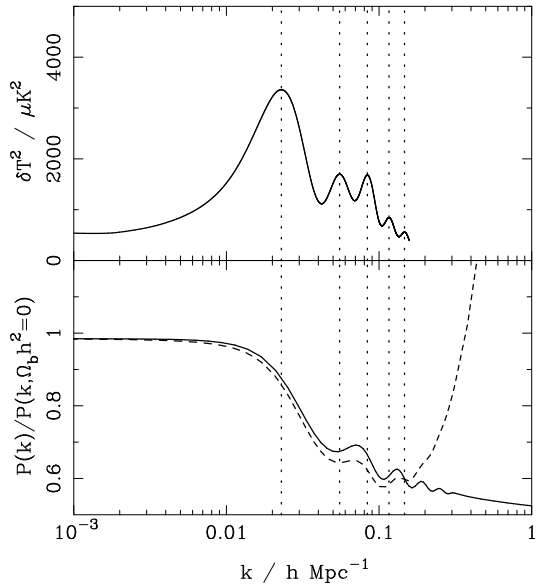}
  \caption{Plot comparing large scale structure (lower panel) and CMB
  (upper panel) power spectra. The angular CMB power spectrum was
  converted to comoving scales using the comoving distance to the last
  scattering surface. The matter power spectrum (solid -- linear,
  dashed -- non-linear, present day), has been ratioed to a smooth
  model with zero baryons in order to highlight the baryonic
  features. Dotted lines show the positions of the peaks in the CMB
  spectrum.}
  \label{fig:lss_cmb}
\end{figure}

Over the past few years there has been a dramatic improvement in the
resolution and accuracy of measurements of fluctuations in the
temperature of the CMB radiation. The discovery of features, in
particular, the first acoustic peak, in the power spectrum of the CMB
temperature has led to a new data-rich era in cosmology
\cite{debernardis00,hanany00}. More recently a significant leap
forward was made with the release of the first year data from the WMAP
satellite \cite{bennett03,hinshaw03}.  The relative positions and
heights of the acoustic peaks encode information about the values of
the fundamental cosmological parameters, as discussed for the matter
power spectrum in Section~\ref{sec:seed}. For a flat cosmological
model with $n_s=1$, $\Omega_M=0.3$, $h=0.7$ and $\Omega_bh^2=0.02$ the
CMB and matter power spectra are compared in
Fig.~\ref{fig:lss_cmb}. In order to create Fig.~\ref{fig:lss_cmb}, the
angular CMB power spectrum was converted to comoving scales by
considering the comoving scale of the fluctuations at the last
scattering surface. In Fig.~\ref{fig:lss_cmb}, the matter power
spectrum has been ratioed to a smooth zero baryon model in in order to
highlight features -- even so, the baryon oscillations are
significantly more visible in the CMB fluctuation spectrum. The
vertical dotted lines in this plot are located at the peaks in the CMB
spectrum and highlight the phase offset between the two spectra. The
CMB peaks are $\pi/2$ out of phase with the matter peaks because they
occur where the velocity is maximum, rather than the density at the
last scattering surface -- this is known as the velocity
overshoot. Additionally there is a projection effect -- the observed
CMB spectrum is the 2D projection of 3D fluctuations, and so is
convolved with an asymmetric function: the projection can increase,
but not decrease the wavelength of a given fluctuation.

\begin{figure}
  \centering
  \includegraphics[width=7cm]{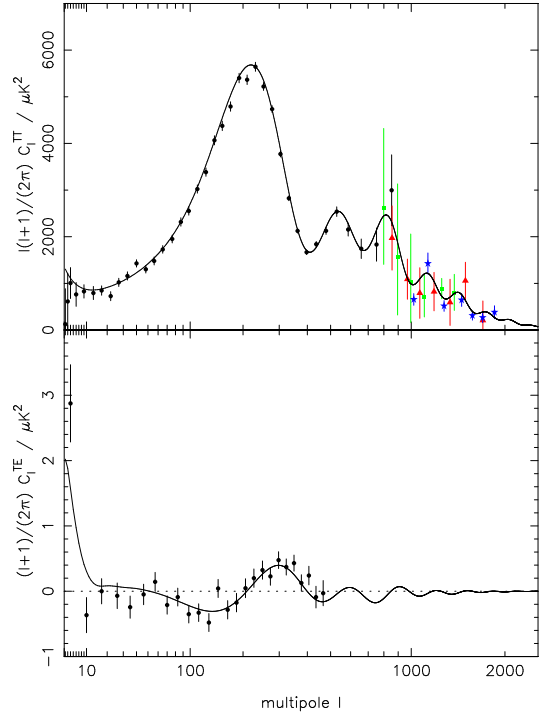}
  \caption{Upper panel: The 1-yr WMAP TT power spectrum (black
  circles) is plotted with the CBI (red triangles), VSA (green
  squares) and ACBAR (blue stars) data at higher $l$. Lower panel: The
  1-yr WMAP TE power spectrum (black circles). In both panels the
  solid black line shows the best fit model calculated from fitting
  the CMB data.}
  \label{fig:cmb_cls}
\end{figure}

A compilation of recent CMB data is presented in
Fig.~\ref{fig:cmb_cls}. Here we have plotted both the
temperature-temperature (TT) auto-power spectrum and the
temperature-E-mode polarisation (TE) cross-power spectrum. The most
significant current data set is, of course, the WMAP data shown by the
solid circles in this figure. However, additional information is
provided on small scales by a number of other experiments. In
Fig.~\ref{fig:cmb_cls}, we plot data from the CBI \cite{cbi}, VSA
\cite{vsa}, and ACBAR \cite{acbar} experiments.

\begin{figure}
  \centering
  \includegraphics[width=\textwidth]{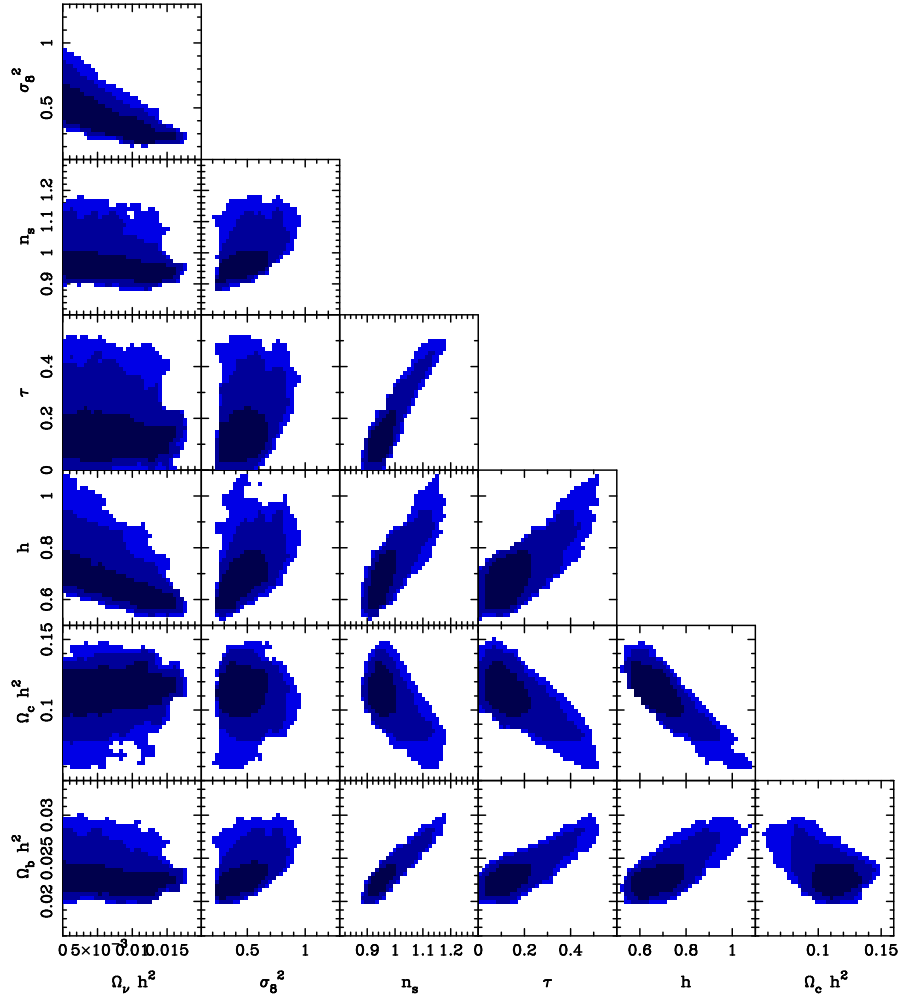}
  \caption{2D projections of the 7D likelihood surface resulting from
  a fit to the CMB data plotted in Fig.~\ref{fig:cmb_cls}. The shading
  represents areas with $-2\Delta{\cal L}=2.3, 6.0, 9.2$ corresponding
  to $1\sigma$, $2\sigma$ and $3\sigma$ confidence intervals for
  multi-parameter Gaussian random variables. There are two primary
  degeneracies - between $\Omega_ch^2$ and $h$ and between $n_s$,
  $\tau$ and $\Omega_bh^2$, which are discussed further in
  Section~\ref{sec:CMB_param_degen}.}
  \label{fig:like_pyramid_cmb}
\end{figure}

Likelihood surfaces from a multi-parameter fit to these CMB data are
shown in Fig.~\ref{fig:like_pyramid_cmb}. For this fit, 7 parameters
were allowed to vary: $\Omega_ch^2$, $\Omega_bh^2$, $h$, $\tau$,
$n_s$, $\sigma_8$, and $\Omega_\nu h^2$. Other cosmological parameters
were set at their ``model simplification'' values as discussed in
Section~\ref{sec:model_param}. In particular, we have assumed a flat
cosmological model with $\Omega_{\rm tot}=1$ and that the tensor
contribution to the CMB is negligible. In choosing this set of $7$
parameters, and using the standard MCMC technique we have implicitly
assumed uniform priors for each. The constraints on the 7 fitted
parameters are given in Table~\ref{tab:param_values}.

\begin{table}
  \centering
  \caption{Summary of cosmological parameter constraints calculated by
  fitting a 7-parameter cosmological model to the CMB data plotted in
  Fig.~\ref{fig:cmb_cls} and to the combination of these data with the
  measurement of the 2dFGRS power spectrum \cite{cole05} -- see text
  for details. Data are given with $1\sigma$ error, except for
  $\Omega_\nu h^2$ which is presented as a 1$\sigma$ upper limit.}
  \label{tab:param_values}
  \begin{tabular}{ccc}
    \hline\noalign{\smallskip}
    parameter & CMB constraint & CMB+2dFGRS constraint \\
    \noalign{\smallskip}\hline\noalign{\smallskip}
    $\Omega_ch^2$ & $0.107\pm0.015$ & $0.106\pm0.006$ \\
    $\Omega_bh^2$ & $0.0238\pm0.0021$ & $0.0235\pm0.00166$ \\ 
    $h$ & $0.725\pm0.096$ & $0.718\pm0.036$ \\ 
    $\tau$ & $<0.204\pm0.117$ & $<0.195\pm0.085$ \\ 
    $n_s$ & $1.00\pm0.064$ & $0.987\pm0.046$ \\
    $\sigma_8$ & $0.703\pm0.125$ & $0.696\pm0.085$ \\  
    $\Omega_\nu h^2$ & $<0.00700$ & $<0.006$ \\
    \noalign{\smallskip}\hline
  \end{tabular}
\end{table}

\subsection{parameter degeneracies in the CMB data}
\label{sec:CMB_param_degen}

\begin{figure}
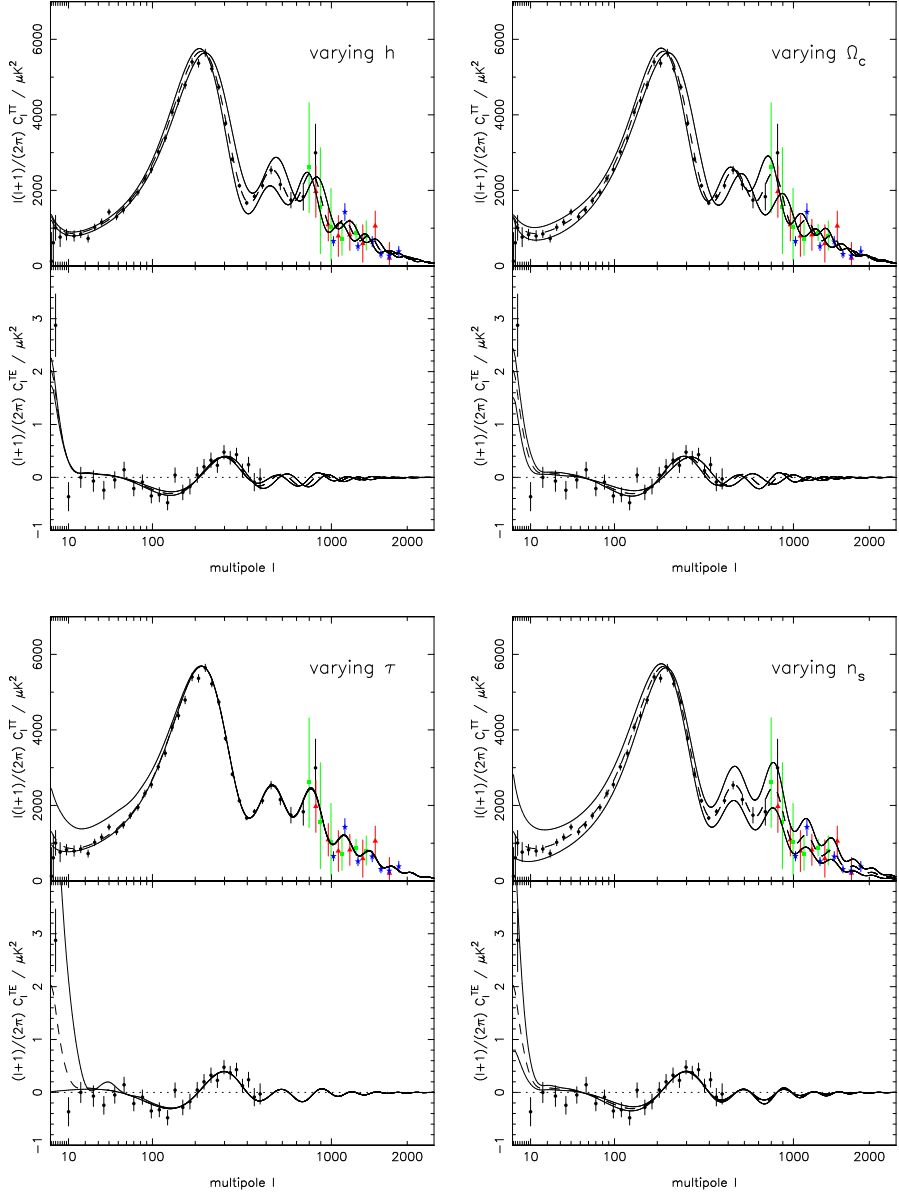

  \centering

  \includegraphics[width=0.48\textwidth]{cmb_cls_vary_h.ps}
  \hfill
  \includegraphics[width=0.48\textwidth]{cmb_cls_vary_omc.ps}

  \vspace{5mm}
  \includegraphics[width=0.48\textwidth]{cmb_cls_vary_tau.ps}
  \hfill
  \includegraphics[width=0.48\textwidth]{cmb_cls_vary_ns.ps}

  \caption{As Fig.~\ref{fig:cmb_cls}, but now showing 3 different
  models: the dashed line shows the best fit model in all panels --
  the model plotted in Fig.~\ref{fig:cmb_cls}. The solid lines in the
  top-left panel were calculated with $h=\pm0.1$, top-right
  $\Omega_c\pm0.1$, bottom-left $\tau+0.3$ and $\tau=0$, and
  bottom-right $n_s\pm0.2$.}
  \label{fig:cmb_cls_vary}
\end{figure}

By examining Fig.~\ref{fig:like_pyramid_cmb} we see that the CMB data
alone do not constrain all of the fundamental cosmological parameters
considered to high precision. Degeneracies exist between certain
combinations of parameters which lead to CMB fluctuation spectra that
cannot be distinguished by current data \cite{efstathiou99}. To help
to explain how these degeneracies arise, CMB models with different
cosmological parameters are plotted in Fig.~\ref{fig:cmb_cls_vary}.

Constraining models to be flat does not fully break the geometrical
degeneracy present when considering models with varying $\Omega_{\rm
tot}$, and a degeneracy between the dark matter density $\Omega_c$ and
the Hubble parameter $h$ remains.  Fig.~\ref{fig:cmb_cls_vary} shows
that both $\Omega_c$ and $h$ affect the location of the first acoustic
peak. A simple argument can be used to show that models with the same
value of $\Omega_mh^{3.4}$ predict the same apparent angle subtended
by the light horizon and therefore the same location for the first
acoustic peak in the TT power spectrum \cite{percival02}. The
degeneracy in Fig.~\ref{fig:like_pyramid_cmb} roughly follows this
prediction.

There is another degeneracy that that can be seen in
Fig.~\ref{fig:like_pyramid_cmb} between $n_s$, $\tau$ and
$\Omega_bh^2$. From Fig.~\ref{fig:cmb_cls_vary}, we see that the
effect of the optical depth $\tau$ on the shape of the TT power
spectrum occurs predominantly at low multipoles. By adjusting the tilt
of the primordial spectrum ($n_s$), the low-$\ell$ power spectrum can
be approximately corrected for the change in $\tau$, and the
high-$\ell$ end can be adjusted by changing the baryon density. This
degeneracy is weakly broken by the TE data which provide an additional
constraint on $\tau$.

\subsection{results from the combination of LSS and CMB data}
\label{sec:final_results}

The CMB degeneracy between $\Omega_c$ and $h$ can be broken by
including additional constraints from the power spectrum of galaxy
clustering. There have been a number of studies using both CMB and
large-scale structure data to set cosmological constraints, with a
seminal paper coming from the WMAP collaboration
\cite{spergel03}. Recently new small-scale CMB data and large-scale
structure analyses have increased the accuracy to which the
cosmological parameters are known. \cite{tegmark04b,sanchez06}.

\begin{figure}
  \centering
  \includegraphics[width=\textwidth]{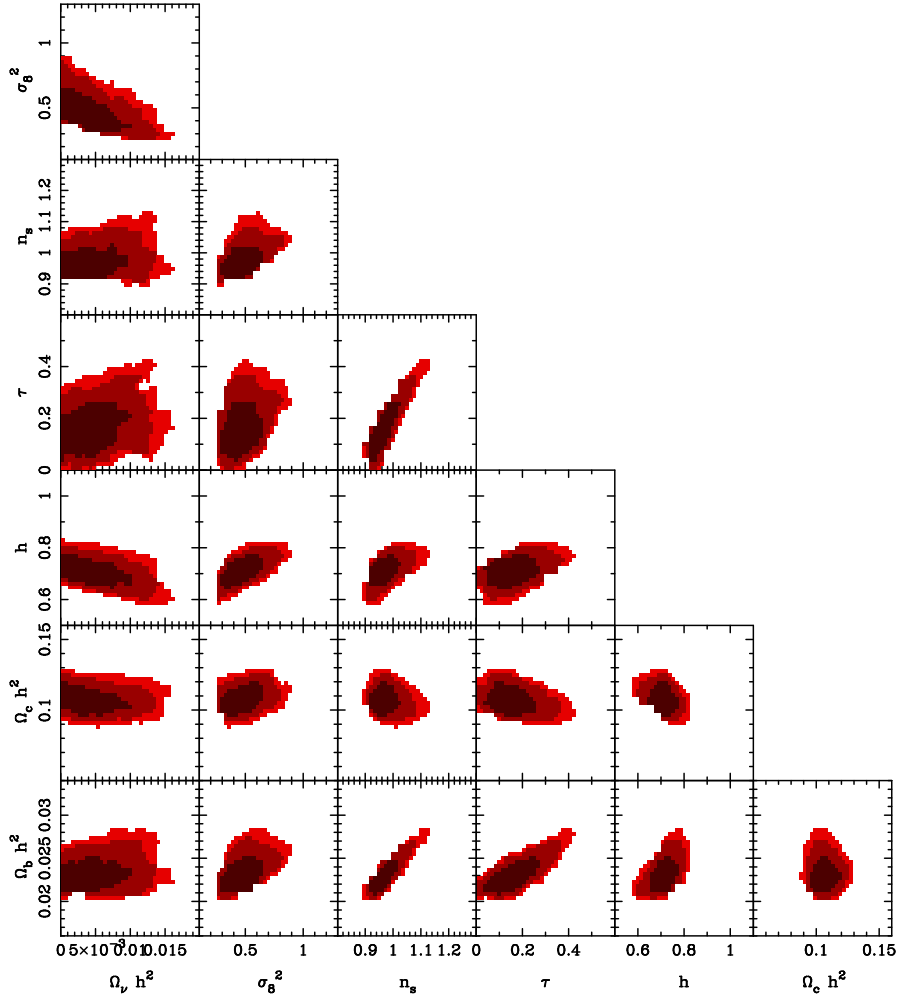}
  \caption{As Fig~\ref{fig:like_pyramid_cmb}, but now including extra
  constraints from the 2dFGRS analysis of \cite{cole05}. These
  constraints helps to break the primary degeneracies discussed in
  Section~\ref{sec:CMB_param_degen}.}
  \label{fig:like_pyramid_cmb_lss}
\end{figure}

In Fig.~\ref{fig:like_pyramid_cmb_lss}, we provide a likelihood plot
as in Fig~\ref{fig:like_pyramid_cmb}, but now including the
cosmological constraints from the final 2dFGRS power spectrum
\cite{cole05}. For this analysis, a constant bias was assumed and we
fitted the galaxy power spectrum over the range
$0.02<k<0.15\hompc$. The derived parameter constraints for the 7
parameters varied are compared with the constraints from fitting the
CMB data only in Table~\ref{tab:param_values}. The physical neutrino
density $\Omega_\nu h^2$ is unconstrained within the prior interval
(physically, it must be $>0$), so we only provide an upper limit.

A Table of parameter constraints, such as that presented in
Table~\ref{tab:param_values} represents the end point of our story. We
have introduced the major steps required to utilise a galaxy survey to
provide cosmological parameter constraints, and have ended up with an
example of a set of constraints for a particular model.



\printindex
\end{document}